\documentclass[pdflatex,sn-mathphys-num]{sn-jnl}
\usepackage{times}
\usepackage{graphicx}%
\usepackage{dsfont}
\usepackage{multicol}
\usepackage[normalem]{ulem}
\usepackage{multirow}%
\usepackage{amsmath,amssymb,amsfonts}%
\usepackage{amsthm}%
\usepackage{mathrsfs}
\usepackage{calrsfs}%
\usepackage[title]{appendix}%
\usepackage[usenames,dvipsnames,x11names]{xcolor}
\usepackage{textcomp}%
\usepackage{manyfoot}%
\usepackage{booktabs}%
\usepackage{algorithm}%
\usepackage{algorithmicx}%
\usepackage{algpseudocode}%
\usepackage{listings}%
\usepackage{graphicx}
\usepackage{subcaption}
\usepackage{lineno}
\usepackage{orcidlink}
\usepackage{float}
\usepackage{pdfpages}
\usepackage{hyperref}
\hypersetup{pdftitle={Editorial}, 
            colorlinks=true,
            citecolor = {cyan},
	        linkcolor = {blue},
            urlcolor  = {purple}
}
\DeclareMathOperator*{\argmin}{\arg\min}
\setcitestyle{comma,open={},close={},super}
\newcommand*{\citen}[1]{%
  \begingroup
    \romannumeral-`\x 
    \setcitestyle{numbers}%
    \cite{#1}%
  \endgroup   
}
\newcommand*{\figref}[1]{%
  \hyperref[{#1}]{%
    Fig.~\ref*{#1}%
  }%
}
\renewcommand*{\eqref}[1]{%
  \hyperref[{#1}]{%
    equation~(\ref*{#1})%
  }%
}
\newcommand*{\tabref}[1]{%
  \hyperref[{#1}]{%
    Table~\ref*{#1}%
  }%
}
\renewcommand*{\algref}[1]{%
  \hyperref[{#1}]{%
    Algorithm~\ref*{#1}%
  }%
}

\author[1]{Sergio Contreras Arredondo\orcidlink{0009-0006-0428-4962}}

\author[1]{Chenyu Tang\orcidlink{0000-0002-6914-7348}}

\author[1]{Radu A. Talmazan\orcidlink{0000-0001-6678-7801}}

\author[2]{Alberto Meg\'ias\orcidlink{0000-0002-7889-1312}}

\author[1]{Cheng Giuseppe Chen\orcidlink{0000-0003-3553-4718}}

\author*[1,3,4]{Christophe Chipot\orcidlink{0000-0002-9122-1698}}
\affil*[1]{Laboratoire International Associ\'e Centre National de la Recherche Scientifique et University of Illinois at Urbana-Champaign, Unit\'e Mixte de Recherche n$^\circ$7019, Universit\'e de Lorraine, B.P. 70239, 54506 Vand\oe uvre-l\`es-Nancy cedex, France}
\affil[2]{Complex Systems Group and Department of Applied Mathematics, Universidad Polit\'ecnica de Madrid, Av. Juan de Herrera 6, E-28040 Madrid, Spain}
\affil[3]{Department of Biochemistry and Molecular Biology,  University of Chicago, Chicago, USA}
\affil[4]{Theoretical and Computational Biophysics Group, Beckman Institute, and Department of Physics, 
University of Illinois at Urbana-Champaign, Urbana, USA}
\email{chipot@illinois.edu}

\title[From Atoms to Dynamics: Learning the Committor Without Collective Variables]{From Atoms to Dynamics: Learning the Committor Without Collective Variables}

\begin{document}
\setlength{\abovedisplayskip}{10pt}
\setlength{\belowdisplayskip}{10pt}
\setlength{\abovedisplayshortskip}{10pt}
\setlength{\belowdisplayshortskip}{10pt} 

\abstract{
This Brief Communication introduces a graph-neural-network architecture built on geometric vector perceptrons to predict the committor function directly from atomic coordinates, bypassing the need for hand-crafted collective variables (CVs). The method offers atom-level interpretability, pinpointing the key atomic players in complex transitions without relying on prior assumptions. Applied across diverse molecular systems, the method accurately infers the committor function and highlights the importance of each heavy atom in the transition mechanism. It also yields precise estimates of the rate constants for the underlying processes. The proposed approach opens new avenues for understanding and modeling complex  dynamics, by enabling CV-free learning and automated identification of physically meaningful reaction coordinates of complex molecular processes.
}

\maketitle

\section*{Introduction}\label{sec:Introduction}
A wide variety of computational methods and strategies have been developed to address rare transitions in complex molecular systems\cite{Peters2016}. Many foundational concepts can be illustrated using a prototypical model featuring two metastable states, $A$ and $B$, within the framework of transition path theory\cite{Vanden-Eijnden-2010}. A particularly important quantity in this context is the committor function, which quantifies the probability that a trajectory starting from any given configuration will reach state $B$ before visiting state $A$\cite{Onsager-1938,bolhuis_reaction_2000}. The committor is often regarded as the optimal one-dimensional reaction coordinate (RC), capturing the progress of the system during the transition between $A$ and $B$\cite{berezhkovskii_one-dimensional_2005,roux2022transition}. Many approaches have been put forth to estimate the committor, from shooting strategies\cite{van_erp_2003,best_2005,peters_2006,lechner,dinner_2005}, 
to the application of a variational principle expressed in terms of its time-correlation function\cite{krivov_2013}. 
More recently, machine learning (ML) methods have emerged as a promising avenue for discovering complex RCs expressed as nonlinear functions of the CVs\cite{chen_2018,Ribeiro2018,bonati_2019}
Furthermore, to enhance sampling of the transition, biasing potentials are  applied to a small set of CVs. These CVs are functions of the atomic positions, and serve as a low-dimensional embedding that preserves the dynamic features of the transition. Said differently, an effective CV should, in general, discriminate between different conformational states and ideally capture the slow modes of the system\cite{prinz_2011,chen_discovering_2023}. The choice of CVs has conventionally been guided by physical and chemical intuition, often relying on predefined descriptors such as interatomic distances, torsional angles, or more complex combinations thereof.
However, this selection based on intuition often lacks quantitative criteria to assess their suitability to encapsulate the correct dynamics\cite{ansari_2022}. Unfortunately, low-dimensional models of the RC are not always sufficient to describe transitions in complex molecular systems with adequate accuracy\cite{prinz_2011}. To address these limitations, empirical methods have recently been replaced by advanced ML approaches, including artificial neural networks (ANNs), which enable a nonlinear combination of large sets of descriptors to define flexible CVs\cite{Bonati_2023}. Still, these methods rely on the selection of appropriate input features that, collectively, contain the essential information about the transition at hand. Lately, much effort has been invested in learning meaningful CVs directly from Cartesian coordinates\cite{ghorbani_2022,zhang_2024,pengmei2025using}, yet often resulting in a substantial increase of the computational cost during molecular simulations. In this work, we propose to bypass the use of explicit CVs altogether by learning the committor function directly from Cartesian coordinates, $\mathbf{x}\equiv \{\mathbf{x}_i\}_{i=1}^N\in \mathbb{R}^{3N}$. Toward this end, we employ graph neural networks (GNN), which can extract relevant geometric and physical information from molecular structures in a data-driven fashion\cite{antib9020012,dror_2020}. GNNs have become a central tool in ML for structured data, enabling the modeling of relationships in systems ranging from social networks to molecules. Early applications in molecular science have leveraged GNNs primarily for their relational reasoning capabilities, treating atoms or residues as nodes, and encoding interactions through edge features, such as distances or chemical bonds\cite{DBLP:journals/corr/abs-1806-01261}. However, early GNN architectures lacked a consistent way to incorporate and transform geometric information, e.g., orientation and spatial symmetry, essential for the accurate description of molecular objects. This limitation has been addressed by the development of equivariant GNNs models that respect symmetry in Euclidean space, recently with the implementation of the more efficient and scalable geometric vector perceptron (GVP) GNNs (GVP-GNNs)\cite{jing2021equivariant}.

The GVP-GNN architecture meets the need for simultaneous geometric and relational reasoning by augmenting message-passing GNNs with GVPs, which operate on both scalar features, $S\in\mathbb{R}^{n}$, and vector features, $\mathbf{V}\in\mathbb{R}^{\nu\times 3}$. These transformations preserve rotation and reflection equivariance on the vector features, while keeping the universal approximation of continuous equivariant functions\cite{jing2021equivariant}. A key innovation in later versions of GVPs lies in the inclusion of vector gating, allowing scalar information to modulate vector channels without breaking equivariance. This makes GVP-GNNs highly expressive, while remaining computationally tractable. This architecture has demonstrated strong empirical performance across diverse structural biology tasks, such as protein-model quality assessment and computational protein design, often outperforming models based on convolutional or purely scalar GNN layers\cite{atom3d}.

The ability of GVP-GNNs to learn functions that are sensitive to spatial structure and symmetry makes them particularly well-suited for computing physically meaningful quantities in molecular systems, such as the committor, which is inherently dependent on the spatial arrangement of atoms. Approaches for estimating the committor that rely on predefined CVs\cite{chen_discovering_2023,MCCCRC25,kang2024Parrinello,jung2023machine} may miss relevant geometric correlations, a shortcoming that can be overcome using the equivariant and relational capacity of GVP-GNNs.

\begin{figure*}[ht]
    \centering
    \includegraphics[width=1.0\textwidth]{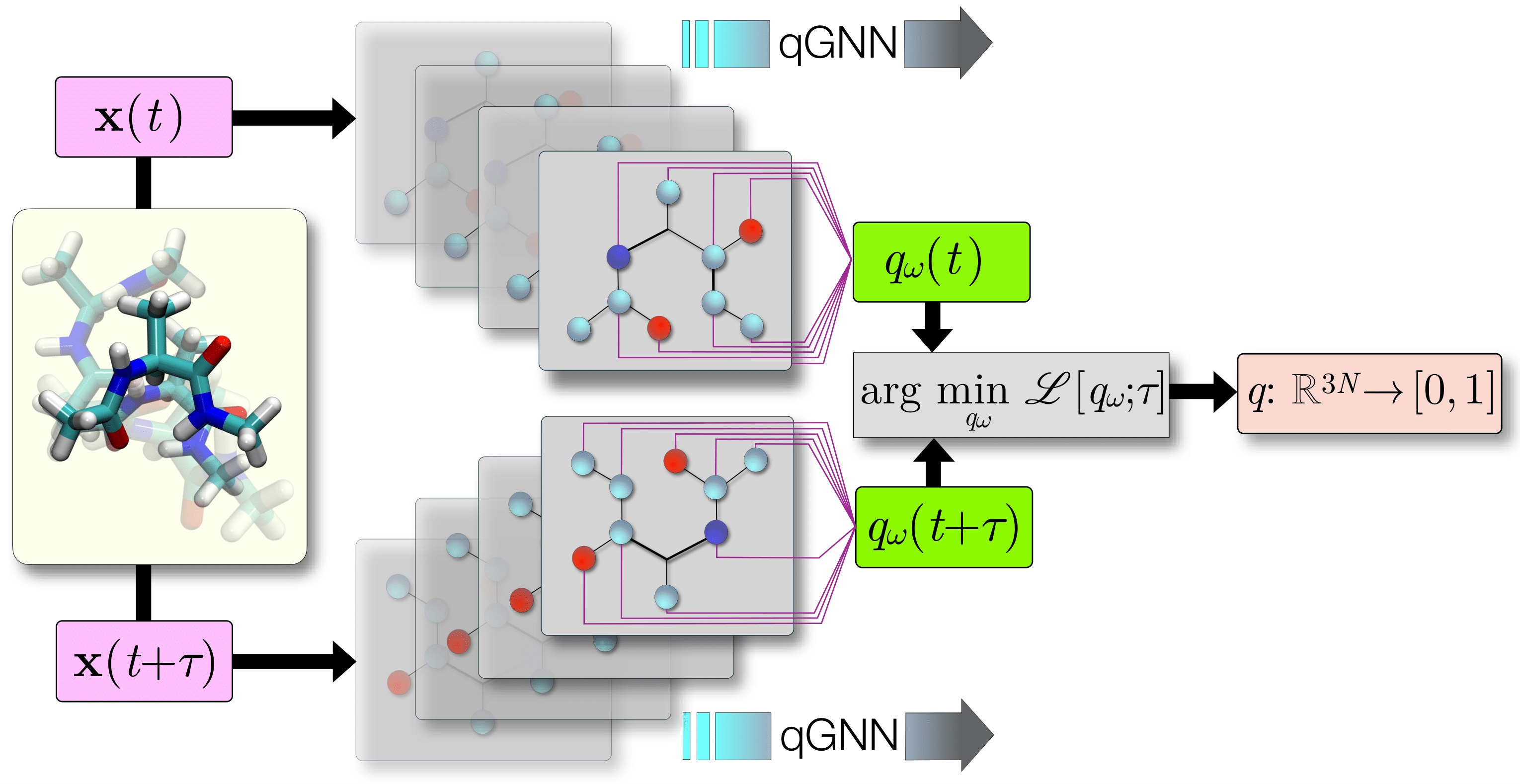}
    \caption{\textbf{qGNN.} Sketch of the learning process for the qGNN. First, we generate a molecular dynamics trajectory---represented by the molecule between brackets---for the studied system and we use this data for training the model. The qGNN model will return a scalar function according to \algref{alg:gnn_learning}. Based on the committor variational principle, we optimize the qGNN scalar output, $q_\omega$, to obtain the committor function, $q$.}
    \label{fig:qGNN}
\end{figure*}

In this Brief Communication, we introduce a methodology for learning the committor, $q$, leveraging the power of GNNs. To achieve this goal, we make use of the well-established committor variational principle\cite{krivov_2013,roux2022transition,chen_discovering_2023,MCCCRC25}, based on the functional optimization of a time-correlation function as explained in \nameref{sec:methods}. This variational principle, as well as the proper boundary conditions of the committor definition (see \nameref{sec:methods}) defines the loss function (see \eqref{eq:loss}) that penalizes the learning of a GVP-GNN model that approximates the committor function in the $\mathbb{R}^{3N}$ space, which we will call the committor-GNN (qGNN). A sketch of the qGNN learning is depicted in \figref{fig:qGNN}. This framework paves the way to the automated, data-driven discovery of RCs in complex molecular systems. We illustrate the generality of our approach across a variety of molecular processes of increasing complexity, including the conformational equilibria of two short peptides, namely $N$-acetyl-$N^{\prime}$-methylalanylamide (NANMA)---also known as dialanine or alanine dipeptide---and a capped trialanine, a Diels–-Alder reaction\cite{ajaz_2011} and the reversible folding of Trp-cage\cite{qiu_2002,Lindorff-Larsen2011fold}. For these different molecular systems, in which the qGNN architecture is maintained identical, we obtain accurate committor predictions based on raw atomic coordinates, and provide meaningful atom-level insights into the RC by means of sensitivity analysis (see \nameref{sec:methods}) as we discuss in \nameref{sec:results}.

\section*{Results}\label{sec:results}
\subsection*{NANMA isomerization}\label{ssec:nanma}
The first application of our methodology is the all-familiar isomerization of NANMA in vacuum, at 300 K. The transition between the C$\mathrm{_{7eq}}$ and C$\mathrm{_{7ax}}$ conformations has been studied previously\cite{chen_discovering_2023,kang2024Parrinello,MCCCRC25}, choosing the backbone dihedral angles, $\phi$ and $\psi$, as the features of the ANNs. 
In stark contrast, use will be made here of the full set of heavy-atom Cartesian coordinates.

\begin{figure}[h!]
    \centering
    \includegraphics[width=0.75\linewidth]{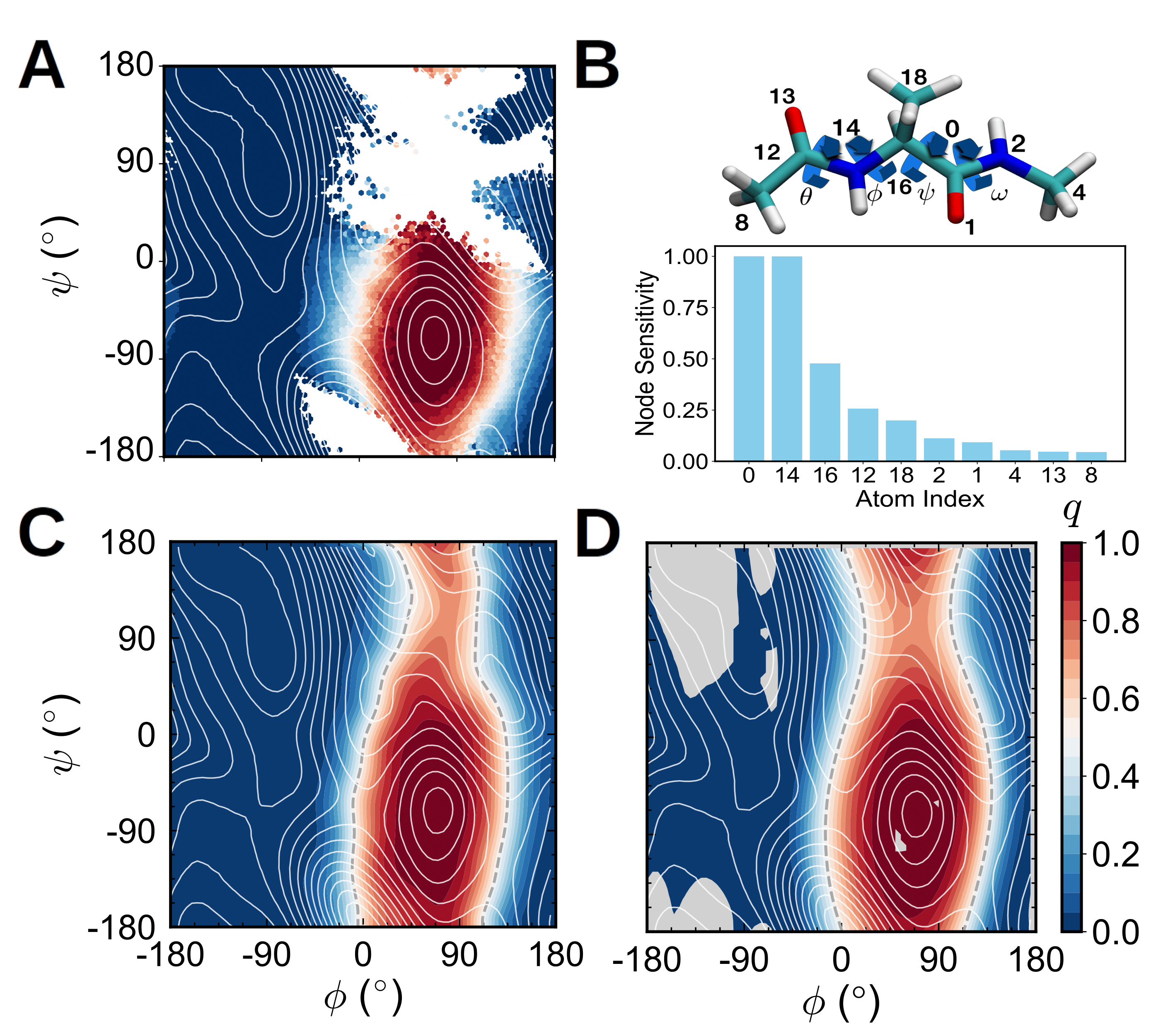}
    \caption{\textbf{NANMA isomerization.} Learned committor $q$, projected on ($\phi$, $\psi$) space, from atomic coordinates of heavy atoms from a biased trajectory of NANMA system. The contour lines represent the free-energy surface (A). Relative node sensitivity, $s_i/\max_j\{ s_j\}$, computed for a short biased simulation of 5ns for the same system (B). Learned committor from the variational committor network\cite{chen_discovering_2023} from a biased trajectory of NANMA sampling the $(\phi, \psi)$ subspace (C) and committor computed from shooting unbiased trajectories for NANMA according to the OpenPathSampling package\cite{SPNCB19,SPNCB19a} (D).}
    \label{fig:nanma}
\end{figure}

Our results are gathered in \figref{fig:nanma}, where the learned committor is projected onto the $(\phi,\psi)$-subspace for visualization purposes. As depicted in \figref{fig:nanma}A, the resulting committor map, learned from a biased simulation (see \nameref{sec:methods}), agrees with previous findings obtained using feed-forward neural networks\cite{chen_discovering_2023,MCCCRC25}, and with numerical results from shooting strategies\cite{SPNCB19,SPNCB19a} shown in \figref{fig:nanma}C and \figref{fig:nanma}D, respectively. The committor map exhibits a well-defined separatrix, i.e., the $q=0.5$ isocommittor hyperplane when projected onto the $(\phi,\psi)$ subspace\cite{chen_discovering_2023,kang2024Parrinello}. Additionally, the node-sensitivity analysis, summarized at \figref{fig:nanma}B, highlights the key heavy atoms that contribute predominantly to the transition process. As expected, our findings confirmed that the most important atoms for the model are the four (14, 0, 16, and 12) that define the dominant dihedral angle, namely $\phi$. Furthermore, the $\psi$ torsional angle, formed by atoms 14, 0, 16 and 2, is also found to be relevant for the transition, as revealed by our analysis.

\subsection*{Trialanine conformational equilibrium}\label{ssec:trialanine}

A similar application was conducted on trialanine. The dihedral angles, $\phi_1$, $\phi_2$, and $\phi_3$, have been reported to be the important CVs to describe the conformational transitions between metastable states $A$ ($\phi_1=60\textdegree$, $\phi_2=-70\textdegree$, $\phi_3=60\textdegree$) and $B$ ($\phi_1=-70\textdegree$, $\phi_2=60\textdegree$, $\phi_3=-70\textdegree$)\cite{chen_companion_2022,tiwary_2016}.  

\begin{figure}[hb]
    \centering
    \includegraphics[width=0.75\linewidth]{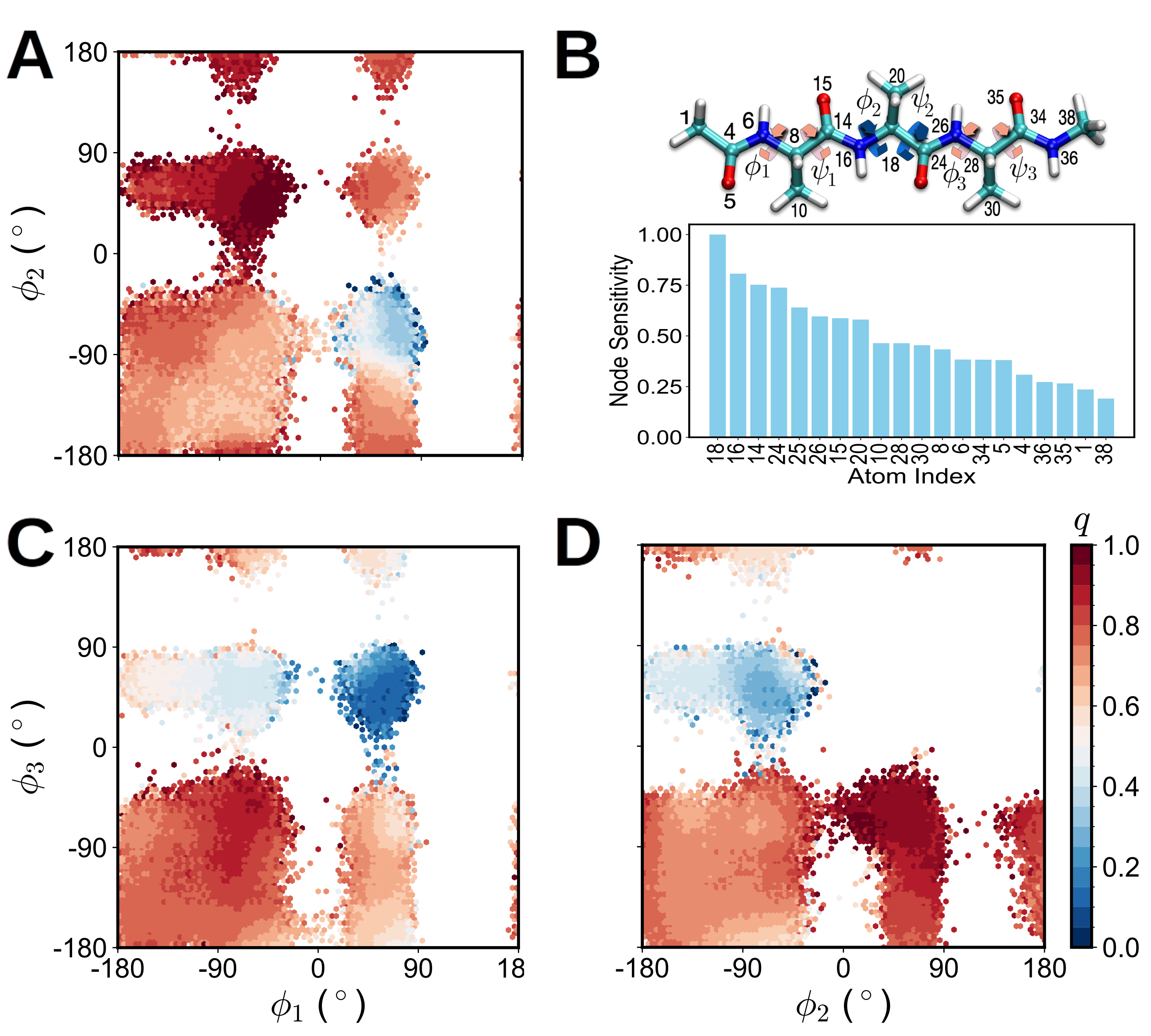}
    \caption{\textbf{Trialanine conformational equilibrium.} Projection of learned committor $q$ onto ($\phi_1$, $\phi_3$) (A), ($\phi_1$, $\phi_2$) (C) and ($\phi_2$, $\phi_3$) (D)  space, from atomic coordinates of a biased trajectory of trialanine. Relative node sensitivity computed for a short biased simulation of 15 ns and a representation of the trialanine molecule (B).}
    \label{fig:triala}
\end{figure}

Our results are presented in \figref{fig:triala}, where the inferred committor is projected onto the $(\phi_1, \phi_2)$-, $(\phi_2, \phi_3)$-, and $(\phi_1, \phi_3)$-subspaces for visualization purposes. The basins associated with both metastable states $A$ and $B$ are clearly identifiable in this representation. The node-sensitivity analysis, shown in \figref{fig:triala}B, highlights the atoms that are most relevant to the transition. Consistent with previous findings\cite{tiwary_2016,chen_companion_2022}, the putative three dominant backbone dihedral angles, $\phi_1$ (atoms 6, 8, 10, and 16), $\phi_2$ (atoms 16, 18, 20, and 26), and $\phi_3$ (atoms 26, 29, 30, and 36), play a central role as confirmed by our analysis. The sensitivity profile supports the notion that these torsional angles are the primary degrees of freedom driving the conformational transition in trialanine.

\subsection*{Diels--Alder reaction}\label{ssec:da}
We have applied the methodology to the reaction of ethylene and but--1--en--3--yne (vinyl--acetylene) to form cyclohexa--1,2--diene, which proceeds via a Diels--Alder mechanism, ubiquitous in organic chemistry. This [4+2] cycloaddition reaction necessitates a quantum-chemical treatment to investigate the creation of the two carbon-carbon bonds. To accelerate  sampling, the semi-empirical GFN2-xTB level of theory was chosen as the quantum-chemical framework, which has proven accurate in Diels--Alder reaction benchmarks\cite{xTB}. A more detailed description of the simulations is available in \nameref{sec:methods}.
\begin{figure}[h!]
    \centering
    \includegraphics[width=0.75\linewidth]{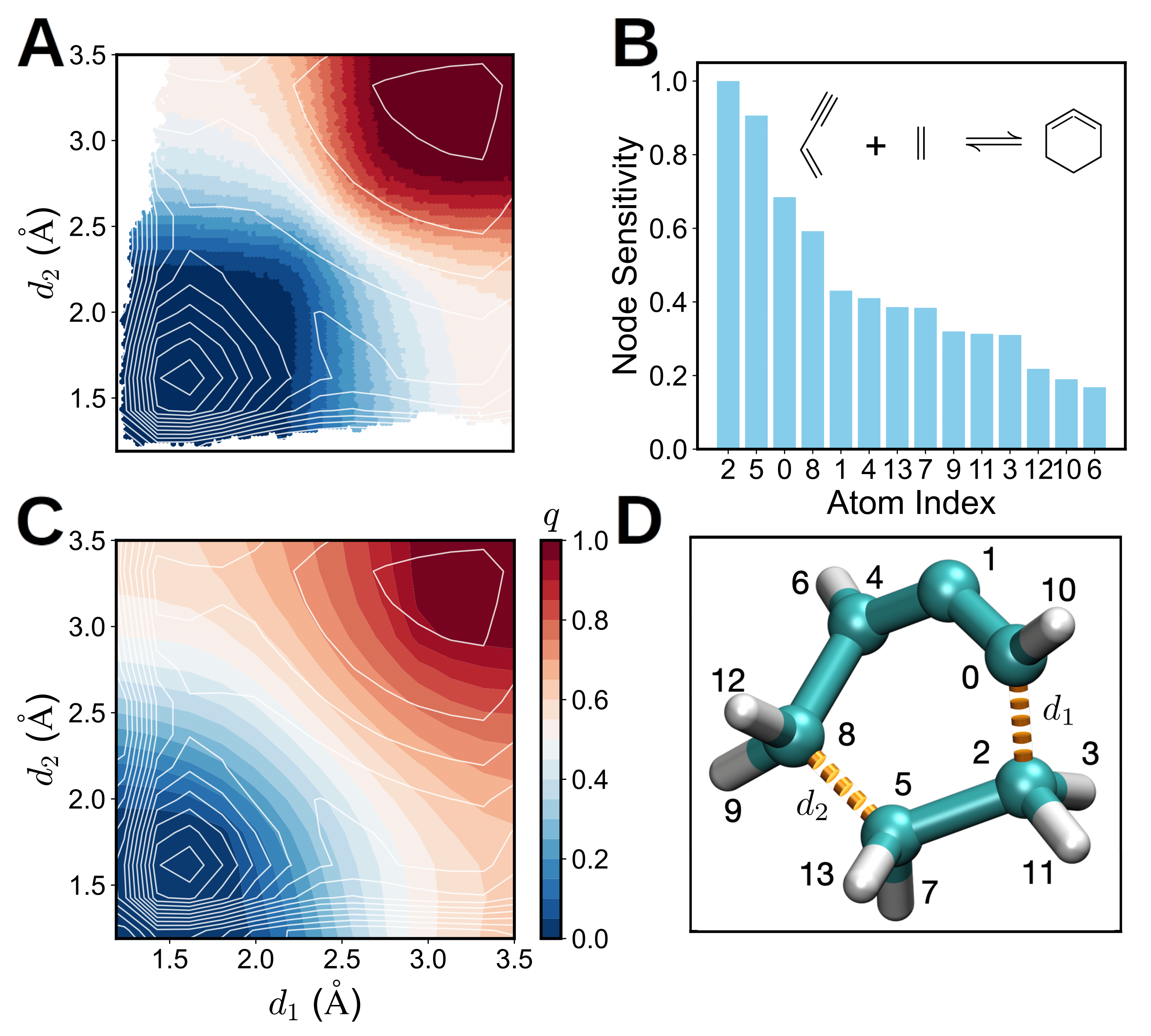}
    \caption{\textbf{Diels--Alder reaction.} Learned committor $q$, projected on ($d_1$, $d_2$)-space, from atomic coordinates of a biased trajectory of Diels--Alder system. The contour lines represent the free-energy surface and the dash-line represents the separatrix (A). Relative node sensitivity analysis computed for a short biased simulation of 2 ns for DA (B). A committor map learned using the variational committor network\cite{chen_discovering_2023} for this reaction is depicted for comparison (C). Furthermore, we show a sketch of the Diels--Alder system with the heavy atoms labelled as well as the distances used in the CV-projection (D).}
    \label{fig:da}
\end{figure}
The initial sampling was performed by biasing the two distances that represent the newly formed bonds, $d_1 = \|\mathbf{x}_0-\mathbf{x}_2\|_2$ and $d_2 = \|\mathbf{x}_5-\mathbf{x}_8\|_2$, where $\|\mathbf{x}_i-\mathbf{x}_j\|_2$ refers to the Euclidean distance between the Cartesian coordinates of atoms $i$-th and $j$-th, represented in \figref{fig:da}D. 

The results in \figref{fig:da}A show the learned committor, with the position of its separatrix in line with earlier theoretical findings\cite{DA}, and in agreement with the results obtained with the CV-based ANN reported in Ref.~\citen{chen_discovering_2023}, as shown in \figref{fig:da}C. The node-sensitivity analysis, the results of which can be found in \figref{fig:da}B, brings to light the four atoms (0, 2, 5, and 8) that participate in the newly created bonds, indicating that the qGNN can discriminate the relevant internal coordinates, namely a pair of distances, amid the Cartesian coordinates. It is noteworthy that atoms 1 and 4 are also found to be significant, forming with atom 0 a 180$^\circ$ angle in the reactant state, which evolves to 120$^\circ$ in the product state.

%

\subsection*{Trp-cage reversible folding}\label{ssec:trp}
As a final application to a realistic biological problem, and to assess the generalizability of our approach, we studied the reversible folding of the 20-residue protein Trp-cage in an explicit aqueous solvent\cite{Lindorff-Larsen2011fold,Meuzelaar2013}.
Toward this end, use was made of molecular dynamics (MD) trajectories kindly provided by D.E. Shaw research\cite{Lindorff-Larsen2011fold}, who conducted long equilibrium unbiased simulations using the Anton supercomputer\cite{Anton,Anton_b}. We refer the reader to the original work for a full detailed description of the simulations. The trajectory spans 208~$\mu$s, and captures multiple folding and unfolding events with atomistic resolution. 

Trp-cage folds within approximately 14~$\mu$s, making it a challenging system to benchmark data-driven approaches for folding mechanisms and committor predictions. Just like with the previous molecular systems, we used the atomic configurations to construct molecular graphs, but, for simplicity, only considered the Cartesian coordinates of the $\mathbf{C}_\mathbf{\alpha}$ atoms, thus providing a coarse-grained, yet informative representation of the protein structure. The qGNN was fed with this subset of atoms to learn the committor. 
\begin{figure*}[ht]
    \centering
    \includegraphics[width=0.92\linewidth]{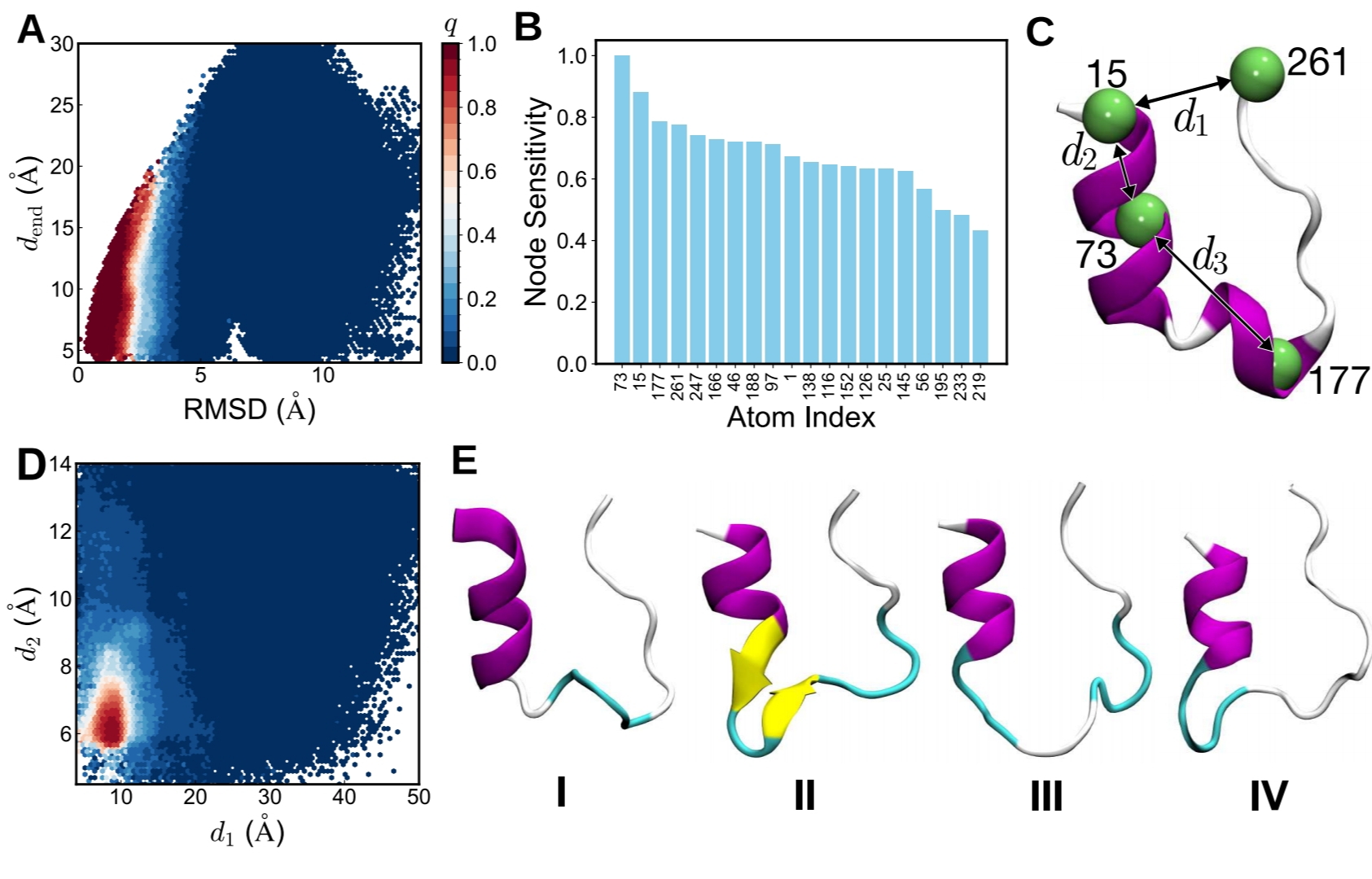}
    \caption{\textbf{Trp-cage reversible folding.} Learned committor $q$, projected on $({\rm RMSD},d_{\rm end})$-space,---where the RMSD is computed with respect to basin $B$---from atomic coordinates of an unbiased trajectory of Trp-cage system. (A). Relative node sensitivity computed for a short unbiased simulation of 50 ns (B). Sketch of Trp-cage protein with the three main distances $d_1$, $d_2$ and $d_3$ (C). The projection of the committor onto the ($d_1$,$d_2$)-subspace (D). Conformations extracted from a clustering analysis in the vicinity of the separatrix, exhibiting distinct secondary-structure elements (E, I to IV).}
    \label{fig:trp}
\end{figure*}
Due to the lack of a reference committor for this system, we initially chose to project the learned committor onto the RMSD and the end-to-end distance, $d_{{\rm end}} \equiv \| \mathbf{x}_{1}-\mathbf{x}_{261}\|_2$, for ease of visualization. As depicted in \figref{fig:trp}A, the committor projected onto these two coordinates reveals a clear separatrix, as well as the two regions associated with the folded  ($q=1$) and unfolded ($q=0$) states. In \figref{fig:trp}B, we present the node-sensitivity analysis, alongside a molecular representation of Trp-cage with the top four most influential atoms highlighted. Based on this information, we computed the edge sensitivity---defined in terms of pairwise distances---among these four most important atoms. The rendered image of the protein also indicates the top three distances ranked by edge sensitivity (see Figure S7 in Section 4 of the SI). While the differences in importance are not as pronounced as in the other systems examined in this Brief Communication, it is worth noting that the first four atoms are involved in three physically meaningful distances, i.e., an end-to-end distance proxy, $d_1\equiv \| {\bf x}_{15}-{\bf x}_{261}\|_2$, a distance related to the extension of the helix, $d_{2} \equiv \| \mathbf{x}_{15}-\mathbf{x}_{73}\|_2$, and an intermediate distance associated with the separation between the two arms of the protein, $d_{3} \equiv \| \mathbf{x}_{73}-\mathbf{x}_{177}\|_2$ (see \figref{fig:trp}C). Based on an edge-sensitivity analysis of the distance features (see Section~4 of the SI), we find that the learned model identifies $d_1$ and $ d_2 $ as the most informative ones for describing the folding process. Accordingly, in \figref{fig:trp}D, we project the committor map onto the $ (d_1, d_2) $-subspace. Additional projections involving $ d_3 $ can be found in Section~4 of the SI.

While the $ (d_1, d_2) $-subspace is expected to be informative based on the sensitivity analysis, the qGNN operates over the full Cartesian-coordinate space. To explore the structural diversity near the transition region, we performed a clustering analysis using the K-means algorithm~\cite{lloyd1982kmeans}, focusing on structures with committor values falling in the $ (0.45, 0.55) $ range, which approximates the separatrix. This analysis, detailed in Section~4 of the SI, revealed four representative structural clusters within the transition ensemble, illustrated in \figref{fig:trp}E. We observe that all the transition states are characterized by a loss of secondary structure with respect to the native folded state, notably in the region encompassing residues 7--16, while the structural features at both ends of the mini-protein are essentially preserved. 

In Table~\ref{table:rate_constant}, we gather all computed rate constants for the different systems, and compare them with published values. As there is no available experimental estimate for the Diels--Alder reaction, we used the Eyring equation\cite{Eyring} to compute this quantity from previously reported coupled-cluster calculations\cite{DA}.

\begin{table}[h!]
        \normalsize
    \begin{tabular}{llll}
        \hline
        \\[-0.2cm]
        & \multicolumn{3}{c}{Transition rate, $k_{AB}$  (ps$^{-1})$} 
        \\[0.1cm] 
        System & GNN & \multicolumn{2}{c}{Reference} 
        \\ 
        \hline
        \\[-0.2cm]
        \textbf{NANMA}      & $3.92\times10^{-5}$ & $1.34\times10^{-5}$ & \citen{chen_discovering_2023}\\
        \textbf{Trialanine} & $5.60\times10^{-5}$ & $3.62\times10^{-5}$ & \citen{chen_companion_2022}$^{\rm a}$\\
        \textbf{Diels--Alder} & $4.40\times10^{-3}$ & $1.03\times10^{-3}$ & \citen{DA}\\
        \textbf{Trp-cage} & $4.67\times10^{-7}$ & $3.2\times10^{-7}$ & \citen{Meuzelaar2013}\\
        & & $2.5\times10^{-7}$ & \citen{qiu_2002}\\
        & & $8.3\times10^{-7}$ & \citen{JB08}\\
        \\[-0.2cm]
        \hline
    \end{tabular}
{\footnotesize $^{\rm a}$ The reported reference value corresponds to the dominant pathway.}

\bigskip

    \caption{{\bf Rate constants.} Summary of the transitions rates, $k_{AB}$, for NANMA and trialanine conformational equilibria, the Diels--Alder reaction, and Trp-cage reversible forlding.
    }
    \label{table:rate_constant}
\end{table}

\section*{Discussion}\label{sec:conclusions}

In this Brief Communication, we introduce a GVP-GNN architecture  designed to learn the committor function directly from atomic coordinates, emancipated from the need of predefined, human-intuited CVs. The resulting qGNN model captures both spatial and relational information relevant for the dynamics of the molecular system at hand. To evaluate the performance of our approach, we trained our model on MD simulations of four prototypical molecular processes of increasing complexity, namely the conformational equilibria of NANMA and trialanine, the formation of cyclohexa–1,2–diene, and the reversible folding of Trp-cage. In all cases, our model produces an accurate and physically meaningful rendering of the underlying dynamics, with committor-derived rate constants consistent with known experimental or theoretical studies. The CV-independence of the qGNN, combined with a clustering analysis, allows transition-ensemble structures to be studied directly in the full Cartesian-coordinate space, as demonstrated in the case of the Trp-cage reversible folding. This approach avoids loss of information associated with CV-based dimensionality reduction. In addition, the node-sensitivity analysis have proven able to discriminate between the most relevant atoms in the transitions, providing valuable information for the definition of physically interpretable CVs, thereby helping improve our  understanding of the underlying molecular mechanisms. Altogether, these results highlight the power and generality of our model to learn complex dynamical features from molecular simulations, offering a scalable and interpretable tool for RC discovery in high-dimensional systems, with an architecture resilient to the complexity of the molecular system. Although qGNN effectively addresses the dependence on predefined CV in committor learning, in contrast to previous works~\cite{jung2023machine,chen_discovering_2023,kang2024Parrinello,MCCCRC25}, we observe that, to some extent, the learning rests upon our choice of the CVs used for biased sampling. Ongoing efforts not only aim at overcoming this limitation, but are also devoted to discovering transition pathways concomitantly with the committor from Cartesian coordinates in the spirit of the recently introduced ANN~\cite{MCCCRC25} for tackling complex molecular processes.

\section*{Methods}\label{sec:methods}

Here, we introduce the theoretical foundations and technical attributes of the qGNN, as well as the computational details of the MD simulations carried out in this work.

\subsection*{qGNN architecture}

The qGNN learning framework builds upon the well-established committor variational principle\cite{krivov_2013,roux2022transition,chen_discovering_2023,MCCCRC25}, based on the functional optimization of a time-correlation function. This variational principle arises from the net forward scalar flux, $J_{AB}$, from a reactant state $A$ to a product $B$, which can be written in the context of TPT as\cite{roux2022transition}
\begin{eqnarray}\label{eq:Cqq}
    J_{AB}[q;\tau]       & = & \frac{C[q;\tau]}{\tau}, 
    \nonumber
    \\
    C[q;\tau] & = & \frac{1}{2}\Big\langle \Big(q(\tau)-q(0)\Big)^{2} \Big\rangle,
\end{eqnarray}
where $\tau$ is the time lag and $q$ fulfills the boundary constraints $q(\mathbf{x})=0$ if $\mathbf{x} \in A$, and $q(\mathbf{x})=1$ if $\mathbf{x} \in B$, consistent with its definition. The committor minimizes $J_{AB}[q;\tau]$ in functional terms satisfying the neccessary condition $\delta J_{AB}[q;\tau]/\delta q=0$, where the choice of $\tau$ guarantees  Markovianity of the transition mechanism. Hence, the time-correlation function $C[q;\tau]$ serves as a loss function in the learning process. A more detailed derivation of the variational principle can be found in Ref.~\citen{roux2022transition} and in the Supplementary Information (SI). 
Hence, given an output of the qGNN, $q_\omega$, we define the following loss function
\begin{align}\label{eq:loss}
    \mathcal{L}[q_\omega;\tau] =& 2C[q_\omega;\tau]+\lambda \mathcal{L}_{AB}[q_\omega],\nonumber \\
    \quad \mathcal{L}_{AB}[q_\omega] =& |q_\omega(\mathbf{x})|^2_{\mathbf{x}\in A}+ |q_\omega(\mathbf{x})-1|^2_{\mathbf{x}\in B},
\end{align}
where $\lambda\in \mathbb{R}^{+}$ is a positive scalar the value of which is selected beforehand, and $\omega$ represents the set of parameters of the qGNN model. Essentially, this loss penalizes the temporal correlation between committor values along both biased and unbiased trajectories\cite{chen_discovering_2023,MCCCRC25}, encouraging the model to find functions that minimize this correlation, while imposing the boundary conditions. Said differently, we optimize the qGNN towards outputting the true committor,
\begin{equation}
    q = \argmin_{q_\omega} \mathcal{L}[q_\omega;\tau].
\end{equation}

Owing to the use of biasing forces in the simulations, corrupting the dynamics of the process at hand, reweighting is necessary\cite{Ribeiro2018,chen_discovering_2023} to determine the unbiased time-correlation functions,
\begin{equation}\label{eq:rew}
    \langle f(\tau)g(0)\rangle = \frac{\displaystyle \langle e^{\delta W(\tau)/2k_{\rm B}T}f(\tau)e^{\delta W(0)/2k_{\rm B}T}g(0)\rangle_b}{\langle \displaystyle e^{\delta W/k_{\rm B}T}\rangle_b},
\end{equation}
where the subscript $b$ stands for the average using the biased ensemble, $\delta W({\bf z})$ is the perturbed potential of mean force (PMF), and, $f$ and $g$ are any functions\cite{BS98}. To guarantee accurate estimates, we only use at the learning stage the part of the simulation that corresponds to a converged free energy.

To compute the committor function using our qGNN model, we follow the workflow summarized in \algref{alg:gnn_learning}, which maps molecular configurations to graph-level predictions. We start by constructing molecular graphs from atomic configurations extracted from MD trajectories. In these graphs, node and edge features encode both spatial and relational information. In other words, from the configuration $\{{\bf x}_i\}_{i=1}^N$ we define a graph $\mathcal{G}$. Next, node and edge features are propagated into a high-dimensional latent space using GVP layers, represented by the operator ${\rm GVP[\cdot]}$, which processes jointly scalar and vector components while respecting geometric equivariance. This operator maps a given pair $(S,\mathbf{V})\in\mathbb{R}^{n}\times \mathbb{R}^{\nu\times 3}$ into another pair in a higher-dimensional subspace, i.e., $(S,\mathbf{V})\in\mathbb{R}^{n}\times \mathbb{R}^{\nu\times 3}\xrightarrow{{\rm GVP}}(S',\mathbf{V}')\in\mathbb{R}^{m}\times \mathbb{R}^{\mu\times 3}$, with $m>n$ and $\mu>\nu$. In the next step, information is propagated through the graph using a message-passing scheme\cite{jing2021equivariant}, where each node updates its representation based on the features of its neighbors and connecting edges. For the last step, to obtain a graph-level prediction that represents the committor of each configuration, we apply a permutation-invariant pooling operation, which, in our implementation, simply consists of a summation over the updated node embeddings. This pipeline enables the model to learn meaningful global features that are then associated with the committor function.

\begin{algorithm}
\caption{Committor computation via qGNN}\label{alg:gnn_learning}
\hspace*{\algorithmicindent} 
\noindent \textbf{Input:} Cartesian coordinates of a given configuration $\mathbf{x}=\{\mathbf{x}_i\}_{i=1}^N$. \\
\noindent \textbf{Output:} Committor value $q$.
\hspace*{\algorithmicindent}  

\begin{algorithmic}[1]
\State $\mathcal{G} \leftarrow (\mathcal{N},\mathcal{E},\mathcal{A})={\rm GNN}[{\bf x}]$
\State $\mathcal{G}' \leftarrow ({\rm{GVP}}[\mathcal{N}],\rm{GVP}[\mathcal{E}],\mathcal{A})$
\State $\mathcal{G'}_{\text{updated}} \leftarrow \text{GVP-GNN}[\mathcal{G'}]$
\State $\mathcal{N}_{q} \leftarrow {\rm{GVP}}[\mathcal{N'}_{\text{updated}}]$
\State \Return $\rm{Pooling}[\mathcal{N}_{q}]$
\end{algorithmic}
\end{algorithm}

It is noteworthy that the architecture of the qGNN proposed herein is identical for all the molecular systems investigated in this Brief Communication. However, the number of nodes of the graph is based on the heavy atoms of the molecular object at hand. In our architecture, we select $n=\nu=1$ for the initial space of tuples, as we are only considering one scalar and one vector feature per node and edge, and $m=2\mu=16$ for the final space of tuples under the GVP action on the nodes and edges. Additionally, we use $\tanh$ as the activation function in the GVP-GNNs, and all the weights in the neural network are initialized according to a Xavier initialization\cite{BG10}. The hyperparameters of the optimization procedure---through an Adam optimizer\cite{kingma_adam_2015}, according to the loss function in \eqref{eq:loss}, are maintained across the different molecular systems, except for the constant $\lambda$ defined in \eqref{eq:loss}. However, even though $\lambda$ is different for each case, a wide range of values of such a parameter leads to comparable results, as explained in the SI. For the optimization procedure according to the variational principle, we define the qGNN in a Siamese-like way, as proposed previously in Refs.~\citen{chen_discovering_2023,MCCCRC25}. Further details on GVP-GNN architectures can be found in Section 1.2. of the SI.

\subsection*{Sensitivity analysis}

To identify and interpret the most relevant structural elements for the learned committor, we performed a node-sensitivity analysis on each system. This analysis reflects the importance of a given node---an atom of the molecular object of interest---for the prediction of the committor value. Therefore, given an MD trajectory of $\ell$ time steps, we can define the graph-trajectory $\mathcal{T} = \{\mathcal{G}_j\}_{j=1}^{\ell}$, where $\mathcal{G}_j \equiv \mathcal{G}(t_j)$ is the graph for the configuration at time $t_j$, $j=1,\dots,\ell$. Next, the absolute sensitivity of the $i$--th element, either node or edge, of the graph related with the atomic structure, $s_i$, is computed as
\begin{equation}
    s_{i}= \frac{1}{\ell}\sum_{j=1}^{\ell} \left \|\left(\frac{\partial q(\mathcal{G})}{\partial \mathbf{V}_{i}} \right)_{\mathcal{G}=\mathcal{G}_j}\right\|_2,
\end{equation}
where $q$ is the committor map obtained from the qGNN, $\mathbf{V}_{i}$ is the vector component of the $i$--th element for which we are computing the sensitivity, and $\| \cdot\|_2$ is the Euclidean norm. From this absolute sensitivity one can define a relative sensitivity as $s_i/\max_j\{ s_j\}$. That is, the subelement of maximum sensitivity will be related to a relative sensitivity of 1, whereas the remaining subelements will take lower values in the interval $[0,1]$. The relative sensitivity of the nodes will be utilized hereafter to examine the importance of each atom in the molecular process of interest.

\subsection*{Simulation details}

\bmhead*{Molecular dynamics simulations.}
In all the studied processes, except for the Trp-cage reversible folding, we performed simulations using the NAMD package\cite{Phillips2020} in conjunction with the Colvars library\cite{Fiorin2024}. The enhanced sampling was generated using the well-tempered metadynamics extended adaptative biasing force  (WTM-eABF) algorithm\cite{fu_taming_2019} on selected CVs with an integration time step of 0.5 fs. Moreover, for all the simulated molecular objects, covalent bonds involving hydrogen atoms were constrained using the RATTLE algorithm\cite{Andersen1983}.

\bmhead*{NANMA isomerization.}

NANMA in vacuum was simulated using the CHARMM22 force field\cite{charmm22}. 100 ns of sampling were generated  using as CVs the RMSD with respect to C$_{\rm 7eq}$ and C$_{\rm 7ax}$, limited to  the heavy atoms (atoms 0, 1, 2, 4, 8, 12, 13, 14, 16, and 18).  The range of the CVs extended from 0.1 $\mathrm{\AA}$ to 1.8 $\mathrm{\AA}$ with a bin width of 0.05 $\mathrm{\AA}$, and the harmonic walls were set at 1.8 $\mathrm{\AA}$ with a force constant of 1 kcal/mol \AA$^2$. Simulations were performed at a constant temperature of 300~K, maintained using a Langevin thermostat with a damping coefficient of 10 ps$^{-1}$. The biasing force  was applied once 10,000 samples were accrued in each bin. The parameters of the extended Langevin dynamics were a damping coefficient of $1~{\rm ps}^{-1}$, an extended fluctuation of $0.02~{\rm \AA}$ and an extended time constant of $100~{\rm fs}$. To avoid sampling of nonphysical regions of large RMSDs, masking of the biasing force was introduced, which was determined beforehand in a 1-ns WTM-eABF simulation. The region was masked for the entire simulation if no sampling was recorded in this prefacing simulation. For the well-tempered metadynamics (WTM) part of the algorithm, the bias temperature was set to 1500~K, with a hill frequency of 1,000,  a hill width of [$0.1~{\rm\AA}$, $0.1~{\rm\AA}$], and a height of 0.1 kcal/mol.

\bmhead*{Trialanine conformational equilibrium.}

Trialanine in vacuum was simulated using the AMBER ff14SB\cite{Maier2015ff14SB}. 100 ns of sampling were gathered from a biased simulation using as CVs the RMSD with respect to metastable states $A$ and $B$, limited to the heavy atoms (atoms 1, 4, 6, 8, 10, 14, 15, 16, 18, 20, 24, 25, 26, 28, 30, 34, 35, 36, and 38). The range of the CVs extended from 0.05 $\mathrm{\AA}$ to 5.05 $\mathrm{\AA}$, with a bin width of 0.1 $\mathrm{\AA}$, and the harmonic walls were set at 5.05 $\mathrm{\AA}$, with a  force constant of 1 kcal/mol \AA$^2$. The protocol of the simulations was identical to that employed for NANMA isomerization, except for the WTM part of the algorithm, for which the bias temperature was set to 1500~K, with a hill width of [$0.2~{\rm\AA}$, $0.2~{\rm\AA}$], and a height of 0.1 kcal/mol.

\bmhead*{Diels--Alder reaction.}  The energies and forces were computed using the external GFN2-xTB quantum-chemical package\cite{xTB}. Restraints were applied on the pair of distances corresponding to the newly formed bonds, in order to limit the sampling of irrelevant structures, whereby the two reactants dissociate completely. A harmonic wall was set at 3.5 $\mathrm{\AA}$, with a 100 kcal/mol \AA$^2$ force constant. To prevent contamination of the sampling with equivalent chemical species having different topologies---e.g., typically a flipped ethylene---a restraint was also applied on the dihedral angle between the two reactants (defined by atoms 6,2,0, and 1). The harmonic walls confining the torsion between $-80^\circ$ and +80$^\circ$ were set with a force constant of 10 kcal/mol degree$^2$. The reaction was simulated with 32 walkers (only four of them where used in the qGNN learning due to memory requirements), sharing gradient information every 1,000 steps, and 2 ns of simulation time per walker. The temperature was set at 300~K using Langevin dynamics, with a damping constant of 10 ps$^{-1}$. The parameters of the extended Langevin dynamics were a damping coefficient of $1~{\rm ps}^{-1}$, an extended fluctuation of $0.01~{\rm \AA}$ and an extended time constant of $200~{\rm fs}$. The sampling along the two distances was partitioned into 73$\times$73 bins, and the biasing force was applied once 5,000 samples were accumulated in each bin. For the WTM part of the algorithm, the metadynamics temperature was set to 2000~K, with a hill frequency of 1,000, a width of $[0.028~{\rm\AA},0.028~{\rm\AA}]$, and a height of 0.1 kcal/mol.

\subsection*{Codes and software}
The GNN committor model designed has been built using the PyTorch Geometric library\cite{pytorchgeometric} and the Geometric Vector Perceptron library\cite{jing2021equivariant}, a rotation-equivariant GNN for learning from biomolecular structure and the programming language Python\cite{python2009}. Trajectories are generated by the popular MD package NAMD\cite{Phillips2020} together with the use of the Colvars module\cite{Fiorin2024}.

\subsection*{Data availability}
All the data in this work can be provided upon request.

\subsection*{Code availability}
All the codes generated in this work can be provided upon request.

\subsection*{Supplementary information}
The article has accompanying supplementary information.

\subsection*{Acknowledgements}
C.C. acknowledges the European Research Council (project 101097272 ``MilliInMicro''), the Université de Lorraine through its Lorraine Université d'Excellence initiative, and the Région Grand-Est (project ``Respire''). C.C. also acknowledges the Agence Nationale de la Recherche under France 2030 (contract ANR-22-PEBB-0009) for support in the context of the MAMABIO project (B-BEST PEPR). A.M. is sincerely thankful for the support provided by Universidad Polit\'ecnica de Madrid through the ``Programa Propio de Investigaci\'on'' grant EST-PDI-25-C0ON11-36-6WBGZ8, as well as the hospitality of Université de Lorraine during his stay. 

\subsection*{Author Contribution Statement}

C.C. conceived the idea, and designed and supervised the research. S.C.A. designed the neural network, developed the corresponding code, and carried out the qGNN learnings. C.T., R.T. and C.G.C. performed the MD simulations. A.M. performed the shooting analysis of NANMA and worked on the theoretical framework. All the authors contributed in the writing of the manuscript and provided helpful discussions.

\subsection*{Competing Interests Statement}

The authors declare no competing interests.

\bigskip




\section*{Supplementary material}

# From Atoms to Dynamics: Learning the Committor Without Collective Variables

# Supplementary Information

# Contents

# 1 Theoretical background

## 1.1 The committor function and the variational principle

The configuration of a molecular system in a specific state is completely determined by Cartesian coordinates $\mathbf{x} \equiv \{\mathbf{x}_i\}_{i=1}^{N} \in \mathbb{R}^{3N}$. The probability density of the system at time $t$ is expressed as $\rho(\mathbf{x}, t)$. The forward propagation step $(\mathbf{x} \to \mathbf{x}')$ for the probability density from the time $t$ to the time $t + \tau$ is

$$\rho(\mathbf{x}, t + \tau) = \int d\mathbf{x} \, \mathscr{P}_\tau(\mathbf{x} \mid \mathbf{x}') \rho(\mathbf{x}', t) \tag{S1}$$

where $\mathscr{P}_\tau(\mathbf{x} \mid \mathbf{x}')$ is the propagator or transfer operator and represents the transition from $\mathbf{x}$ to $\mathbf{x}'$. The dynamics of the system is assumed to be Markovian with a finite time-lag $\tau$, and $\mathscr{P}_\tau$ obeys the Chapman-Kolmogorov equation $\rho(t + n\tau) = \mathscr{P}_{n\tau} \rho(t)$, with $\mathscr{P}_{n\tau} = (\mathscr{P}_\tau)^n$. It is assumed that the system is in thermodynamic equilibrium and that we have microscopic

1

detailed balance, $\mathscr{P}_\tau(\mathbf{x} \mid \mathbf{x}')\rho_{\text{eq}}(\mathbf{x}') = \mathscr{P}_\tau(\mathbf{x}' \mid \mathbf{x})\rho_{\text{eq}}(\mathbf{x})$. Let us assume a system with two metastable states $A$ and $B$, the forward committor, $q(\mathbf{x})$, is the sum of the probability over all paths starting at $\mathbf{x}$ that ultimately reach the states $B$ before ever reach the state $A$, by definition. The probability of each of these paths is expressed as a product of discrete propagation steps $\mathscr{P}_\tau \cdots \mathscr{P}_{n\tau}$ with time-lag $\tau$, under the restriction that the intermediate states resulting from all of these steps are in $A^c \cap B^c$. Summing over all possible paths, it follows that $q(\mathbf{x})$ can be written as

$$q(\mathbf{x}) = \int \mathrm{d}\mathbf{x}' \ q(\mathbf{x}')\mathscr{P}_\tau(\mathbf{x}'|\mathbf{x}) \tag{S2}$$

with the constraints $q(\mathbf{x}) = 0$ if $\mathbf{x} \in A$, and $q(\mathbf{x}) = 1$ if $\mathbf{x} \in B$, arising from its definition. It should be noticed that the equation for the committor probabilities involving only the elementary propagator $\mathscr{P}_\tau(\mathbf{x}'|\mathbf{x})$ for the time lag, is valid under Markovianity of the dynamics as expressed by the Chapman-Kolmogorov equation. In the context of TPT, one can express the net forward reactive flux from $A$ to $B$ as

$$J_{AB} = \frac{1}{2\tau} \int \mathrm{d}\mathbf{x} \int \mathrm{d}\mathbf{x}' \left(q(\mathbf{x}') - q(\mathbf{x})\right)^2 \mathscr{P}_\tau(\mathbf{x}'|\mathbf{x})\rho_{\text{eq}}(\mathbf{x}), \tag{S3}$$

which, equivalently, can also be expressed in terms of the time-correlation functional[1] of equation (1) in the main text. The expression for $J_{AB}$ can serve as a robust variational principle to optimize a trial committor. Minimizing the quantity $J_{AB}$ in the functional sense with respect to a trial function $q$, $\delta J_{AB}/\delta q = 0$ recovers equation (S2) that formally defines the committor probability. Thus, defining the committor time-correlation functional $C[q;\tau] = \langle(q(\tau) - q(0))^2\rangle/2$, one can use the loss function to refine a trial model function for the committor probability[2,3]. A more detailed discussion about the variational principle can also be found in Ref. [1].

## 1.2 GVP-GNNs architectures

Learning from macromolecular structures in three-dimensional space requires neural architectures that are both rotation-equivariant and expressive in terms of relational reasoning. Traditional GNNs are effective at modeling relational structures, whereas convolutional neural networks handle the geometric information of the systems. Geometric Vector Perceptrons (GVPs) were introduced as a unified method that bridges graph neural networks (GNNs) and convolutional neural networks to simultaneously capture the full geometric and relational information encoded in macromolecular systems, building GVP-GNNs as a new class of GNNs that extend scalar-based message-passing schemes to operate on both scalar and geometric vector features.

The GVP learns functions which output vectors or scalars, using geometric vectors and scalars as inputs as explained in the main text, i.e., given a tuple $(S, \mathbf{V})$ of scalar features $S \equiv \{s_i\}_{i=1}^n \in \mathbb{R}^n$ and vector features $\mathbf{V} \equiv \{\mathbf{v}_i\}_{i=1}^\nu \in \mathbb{R}^{3\times\nu}$—where each $\mathbf{v}_i$ belongs to the three-dimensional space,—we compute new features $(S', \mathbf{V}') \in \mathbb{R}^m \times \mathbb{R}^{3\times\mu}$. In the architecture proposed here, we consider graphs as representations of atomic-level structures in which all node and edge embeddings are tuples $(S, \mathbf{V})$, as detailed in the main text. Information is propagated through the network layers using a message-passing scheme[4–6], where

node features are updated based on information from their neighbors. GVP-GNNs utilize this mechanism by aggregating information from neighboring nodes and edges to iteratively update both scalar and vector node embeddings during each propagation step. We briefly outline the two fundamental components of our model, the GVP layer and the overall GVP-GNN architecture, built by integrating GVPs with message-passing. For full implementation details, we refer the reader to the original works in Ref. 4, 7. Considering that a GVP layer operates on tuples $(S, \mathbf{V})$, this transformation is defined in Algorithm S1, as reported in Ref. 7.

---

**Algorithm S1** Operation of GVP layers[7]

---

**Input:** Tuple (scalar and vector features) $(S, \mathbf{V})$.
**Output:** Tuple (scalar and vector features) $(S', \mathbf{V}')$.
**GVP**:
1: $\mathbf{V}_h \leftarrow \mathbf{W}_h \mathbf{V}$
2: $\mathbf{V}_\mu \leftarrow \mathbf{W}_\mu \mathbf{V}_h$
3: $S_h \leftarrow \|\mathbf{V}_h\|_2$
4: $S_{h+n} \leftarrow \text{concat}(S_h, S)$
5: $S_m \leftarrow \mathbf{W}_m S_{h+n} + \mathbf{b}_m$
6: $S' \leftarrow \sigma(S_m)$
7: $\mathbf{V}' \leftarrow \sigma_g(\mathbf{W}_g[\sigma^+(S_m)] + \mathbf{b}) \odot \mathbf{V}_\mu$
8: **return** $(S', \mathbf{V}')$

---

In Algorithm S1, we have introduced $\sigma$ and $\sigma^+$, which are scalar nonlinearities (e.g., ReLU and identity), and $\sigma_g$ is a gating function (e.g., sigmoid). The operation $\odot$ denotes row-wise multiplication, where each 3D vector in a matrix is scaled by a corresponding scalar, while $\mathbf{W}_h$, $\mathbf{W}_\mu$, $\mathbf{W}_m$ and $\mathbf{W}_g$ are linear learnable transformations, $\mathbf{b}_m$, $\mathbf{b}$ are learnable biases and $\|\cdot\|_2$ is the Euclidean norm. This formulation ensures that the vector and scalar outputs of the GVP are equivariant and invariant, respectively, with respect to an arbitrary composition of rotations and reflections in 3D Euclidean space. Formally, invariance and equivariance can be defined as follows,

**Definition 1.** *Let $G$ be a group. A map $f : X \to Y$ is called invariant with respect to the $G$-action $\cdot$ on $X$ and $Y$, if*

$$f(g \cdot x) = f(x), \quad \forall x \in X, \forall g \in G.$$

**Definition 2.** *Let $G$ be a group. A map $f : X \to Y$ is called equivariant with respect to the $G$-action $\cdot$ on $X$ and $Y$, if*

$$f(g \cdot x) = g \cdot f(x), \quad \forall x \in X, \forall g \in G.$$

As an example, if $R$ is a composition of rotations, and if $\text{GVP}(S, \mathbf{V}) = (S', \mathbf{V}')$, then $\text{GVP}(S, R(\mathbf{V})) = (S', R(\mathbf{V}'))$. This equality stems from the fact that the only permissible operations on vector inputs are scalar scaling, linear combinations, and the Euclidean norm. In addition, the GVP architecture inherits the Universal Approximation Theorem of dense

layers with respect to rotation and reflection equivariant functions as stated by the theorem in Refs. [4, 7], that reads as follows.

**Theorem 1.** *Let $R$ describe an arbitrary rotation and/or reflection in $\mathbb{R}^3$ and $G_s$ be the vector outputs of a GVP defined with $n = 0$, $\mu = 6$, and sigmoidals $\sigma$, $\sigma^+$. For $\nu \geq 3$, let $\Omega^\nu \subset \mathbb{R}^{\nu \times 3}$ be the set of all $\mathbf{V} = [\mathbf{v}_1, \ldots, \mathbf{v}_\nu]^\top \in \mathbb{R}^{\nu \times 3}$ such that $\mathbf{v}_1, \mathbf{v}_2, \mathbf{v}_3$ are linearly independent and $0 \leq \|\mathbf{v}_i\|_2 \leq b$ for all $i$ and for some $0 < b < \infty$. Then, for any continuous and rotation/reflection-equivariant $F : \Omega^\nu \to \mathbb{R}^3$, and for any $\varepsilon > 0$, there exists a form $f(\mathbf{V}) = \mathbf{1}^\top G_s(\mathbf{V})$ such that $|F(\mathbf{V})_i - f(\mathbf{V})_i| < 6bC\varepsilon$, with $0 < C < \infty, \forall i \in \{1,2,3\}$ and $\forall \mathbf{V} \in \Omega^\nu$.*

Additionally, this theorem can be generalized to $n \neq 0$, that is, over functions defined on $\mathbb{R}^n \times \mathbb{R}^{\nu \times 3}$.[7]

The full GVP-GNN architecture is built by combining GVPs with message-passing.[4] Each node in the graph carries scalar and vector features. Messages passed from node $j$ to node $i$ are computed via

$$\mathbf{h}_m^{(j \to i)} = g\left(\text{concat}(\mathbf{h}_v^{(j)}, \mathbf{h}_e^{(j \to i)})\right)$$
$$\mathbf{h}_v^{(i)} \leftarrow \text{LayerNorm}\left(\mathbf{h}_v^{(i)} + \frac{1}{k'} \sum_{j: e_{j \to i} \in E} \text{Dropout}(\mathbf{h}_m^{(j \to i)})\right) \tag{S4}$$

where $g$ is a stack of GVPs, $k'$ is the number of neighbors (i.e. incoming messages), $\text{concat}(\cdot, \cdot)$ is the concatenation function, $\text{Dropout}(\cdot)$ is a regularization scheme that helps neural networks to be prevented from overfitting[8] and $\text{LayerNorm}(\cdot)$ is the layer normalization. The $\mathbf{h}_v^{(j)}$ and $\mathbf{h}_e^{(j \to i)}$ stand for nodes and edges embeddings, respectively, while $\mathbf{h}_m^{(j \to i)}$ describes the message from node $j$ to node $i$. A feedforward step updates node embeddings as,

$$\mathbf{h}_v^{(i)} \leftarrow \text{LayerNorm}\left(\mathbf{h}_v^{(i)} + \text{Dropout}(g(\mathbf{h}_v^{(i)}))\right). \tag{S5}$$

This formulation allows for both scalar and geometric vector features to evolve jointly in an equivariant fashion across graph layers. Finally, since we are primarily interested in learning a global feature, namely, the committor function—that characterizes the state of the entire graph—it is necessary to aggregate information from all nodes and edges into graph-level quantities. This is typically achieved using permutation-invariant pooling functions, such as a simple summation or averaging over the output scalar node features of the network[9].

## 2 Transition rates

For all the illustrations reported here, we assume a two-state Markov jump between metastable states $A$ and $B$,

$$A \underset{k_{BA}}{\overset{k_{AB}}{\rightleftharpoons}} B \tag{S6}$$

from which a transition rate can be determined. The equilibrium state probabilities are $p_A = k_{BA}/(k_{AB} + k_{BA})$ and $p_B = k_{AB}/(k_{AB} + k_{BA})$. The unidirectional fluxes are equal at equilibrium, that is, $J_{AB} = p_A k_{AB} = J_{BA} = p_B k_{BA}$. The mean first passage time (MFPT) from $A$ to $B$ can be determined as $1/k_{AB}$.[1,10,11] In our implementation, in the spirit of the VCN,[2] the transition rate can be obtained from $A$ to $B$ from the unidirectional flux $k_{AB} =$

$J_{AB}/p_A$, where the flux $J_{AB}$ is given by Equation 1 of the main manuscript in terms of the time-correlation function of the committor. Here, we determine the transition rate, $k_{AB}$, from the slope of the time-correlation function $C_{qq}(\tau)$, as a function of $\tau$ and $p_A$, for the different systems examined.

## 2.1 NANMA isomerization

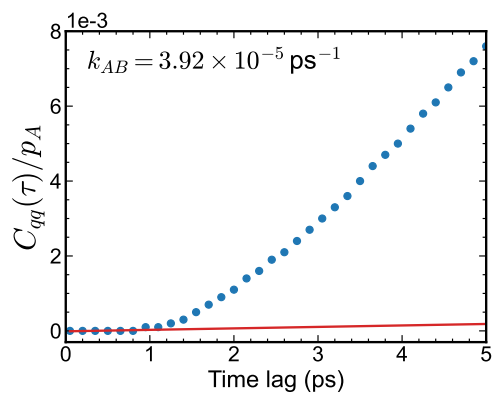

**Fig. S1**: **Fitting of the rate constant for NANMA.** Rate constant $k_{AB}$ determined from GNN committor using the biased trajectory of NANMA.

## 2.2 Trialanine conformational equilibrium

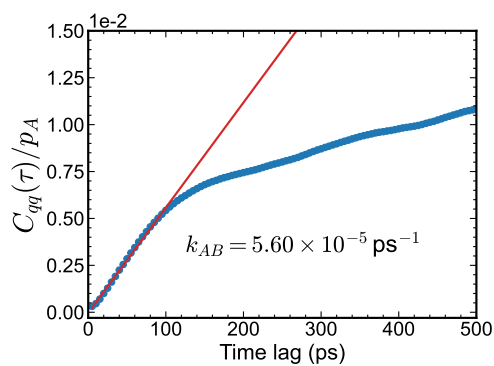

**Fig. S2**: **Fitting of the rate constant for trialanine.** Rate constant $k_{AB}$ determined from GNN committor using the biased trajectory of trialanine.

## 2.3 Diels–Alder reaction

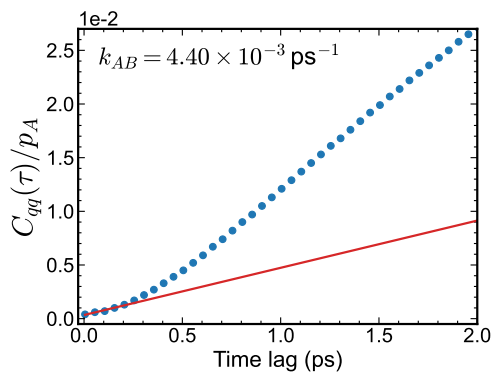

**Fig. S3**: **Fitting of the rate constant for Diels–Alder reaction.** Rate constant $k_{AB}$ determined from GNN committor using the biased trajectory for the Diels–Alder reaction.

## 2.4 Trp-cage reversible folding

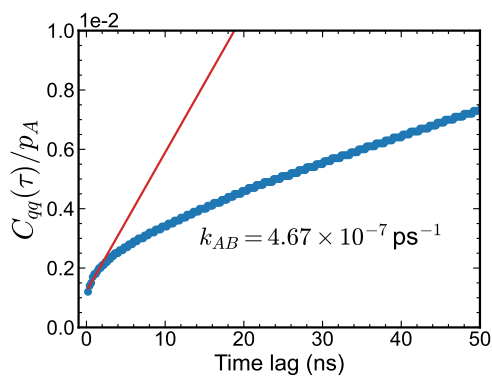

**Fig. S4**: **Fitting of the rate constant for Trp-cage reversible folding.** Rate constant $k_{AB}$ determined from GNN committor using the biased trajectory for the Trp-cage.

## 3 Dependence on the restraints applied to the basins ($\lambda$)

Here, in Fig. S5–S6, we show how learning differs with different choices of the parameter $\lambda$ of the training loss function (see equation (2) of the main paper). Only in those cases where $\lambda$ is exaggerated to values of the order of $10^5$, we encounter either inaccurate or incomplete learning. However, through a wide range of values the output is consistent.

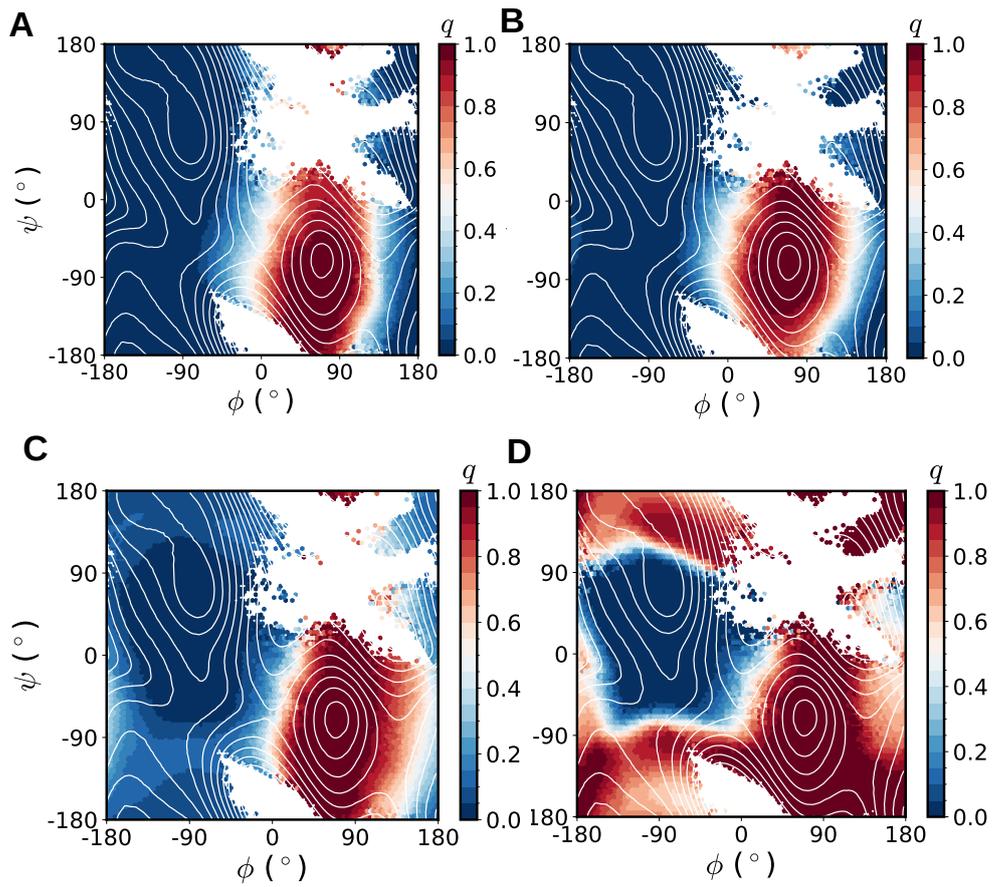

**Fig. S5**: **Learned committor for NANMA and different values of** $\lambda$. Learned committor projected onto $(\phi, \psi)$ subspace for $\lambda = 0.5$ (A), $\lambda = 10$ (B), $\lambda = 100$ (C) and an extreme value of $\lambda = 10000$ (C).

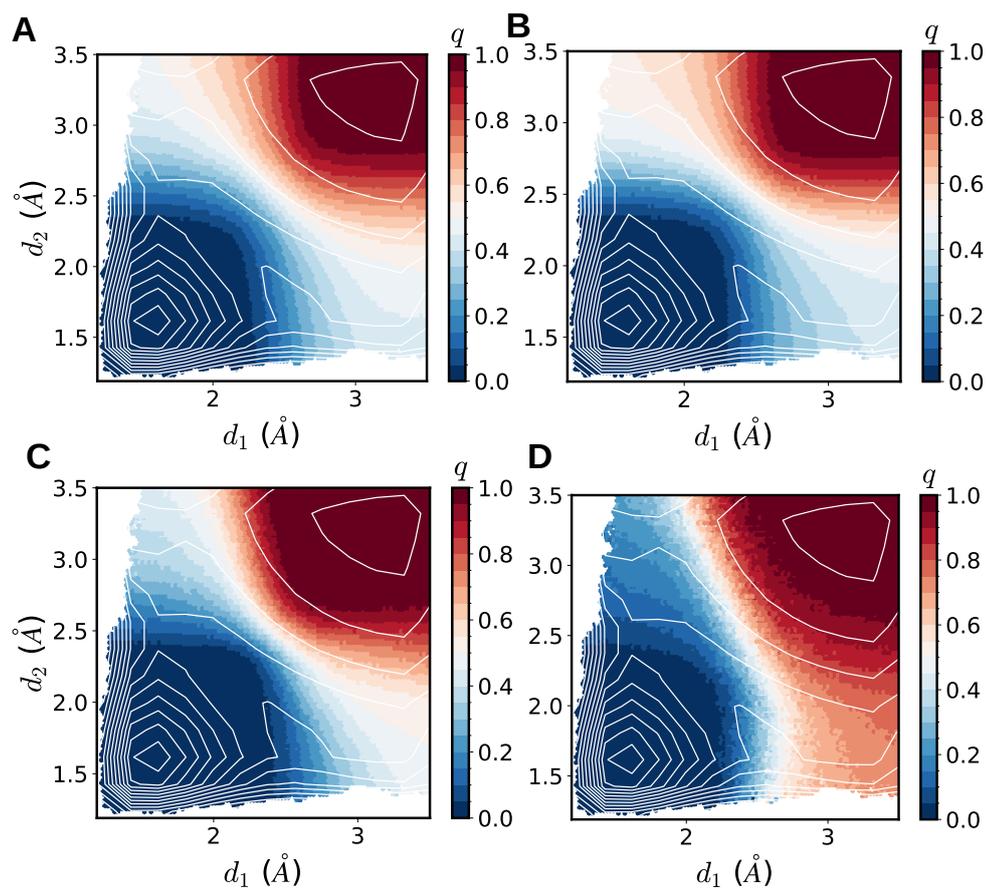

**Fig. S6**: **Learned committor for Diels–Alder reaction and different values of** $\lambda$. Learned committor projected onto $(d_1, d_2)$ subspace for $\lambda = 0.5$ (A), $\lambda = 10$ (B), $\lambda = 100$ (C) and extreme values of $\lambda = 10000$ (C).

# 4 Trp-cage system

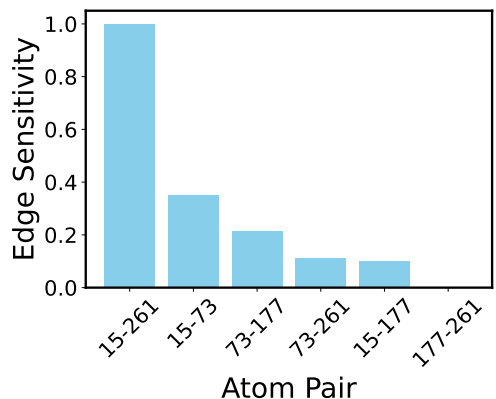

**Fig. S7**: **Edge-sensitivity of distances in Trp-cage from the unbiased simulation.** Edge sensitivity computed from a short biased simulation of 10 ns.

In addition to the unbiased simulation taken from the previous study on the fast-folding protein Trp-cage [12], we also tested the performance of our model on a WTM-eABF enhanced sampling of the folding dynamics of the Trp cage protein using the CHARMM36m force field. [13] To do so, we performed a 10-ns biased simulation along the RMSD with respect to the folded structure limited to the $C_\alpha$ atoms. (PDB: 2JOF) [14] The range of the CV extended from 0.0 Å to 10.0 Å with a bin width of 0.2 Å, and a harmonic wall was set at 10.5 Å with a force constant of 100 kcal/mol Å$^2$. The simulation was maintained using a Langevin thermostat with a damping coefficient of 10 ps$^{-1}$ at a constant temperature of 300 K. The biasing force was applied once 10,000 samples were collected in each bin. The parameters of the extended Langevin dynamics were a damping coefficient of 1 ps$^{-1}$, an extended fluctuation of 0.1 Å and an extended time constant of 100 fs. For the well-tempered metadynamics part of the algorithm, the bias temperature was set to 1000 K, with a hill frequency of 1,000, a hill width of [0.2 Å, 0.2 Å], and a height of 0.1 kcal/mol.

Our results for the short-biased simulation and long-unbiased simulations are generally similar. However, the differences in node sensitivity may indicate that adding CV-dependent biases during the simulations can influence the model's prediction of the primary degrees of freedom. The reader is, therefore, advised to proceed with care when running biased simulations.

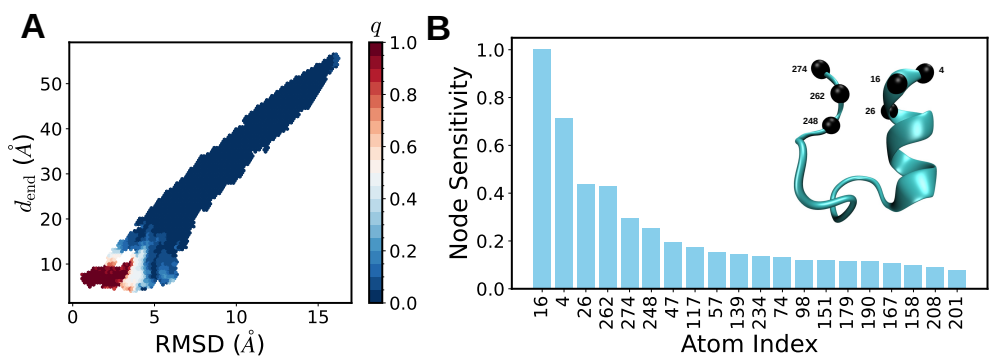

**Fig. S8**: **Trp-cage committor from a biased trajectory.** Learned committor from the biased simulation projected onto RMSD and end-to-end distance, from atomics coordinates of a biased trajectory of Trp-cage system (A). Node sensitivity computed from a short biased simulation fo 10 ns (B).

**Committor projections onto the $(d_1, d_3)$- and $(d_2, d_3)$-subspaces.** In Fig. S9, one can observe the committor projections onto the $(d_1, d_3)$- and $(d_2, d_3)$-subspaces to complete the scheme provided by the first three main distances in the edge-sensitivity analysis.

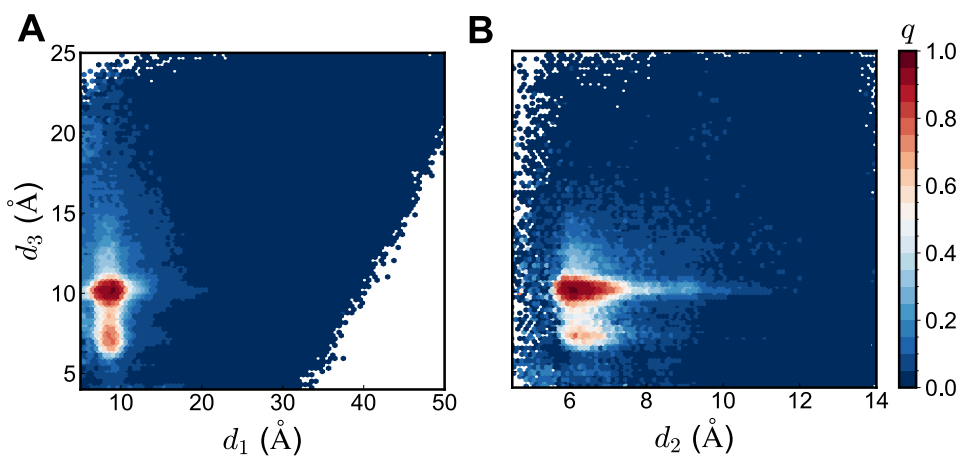

**Fig. S9**: **Role of $d_3$ in the committor of Trp-cage reversible folding from the unbiased simulation.** Projection of the learned committor using unbiased trajectory[12] onto the $(d_1, d_3)$ (A) and $(d_2, d_3)$-subspaces (B).

**Clustering at the separatrix.** A clustering of the simulation frames in the vicinity of the separatrix, i.e. with a committor value between 0.45 and 0.55, was performed, using the K-means algorithm, as implemented in the sklearn python package[15]. By analyzing the clustering quality via the silhouette,[16] Calinski-Harabazs,[17] and Davies-Bouldin scores,[18] we

10

observed, as shown in Fig. S10, an elbow in the values as the cluster number increases. Thus, we determined that 4 represents a suitable number of clusters.

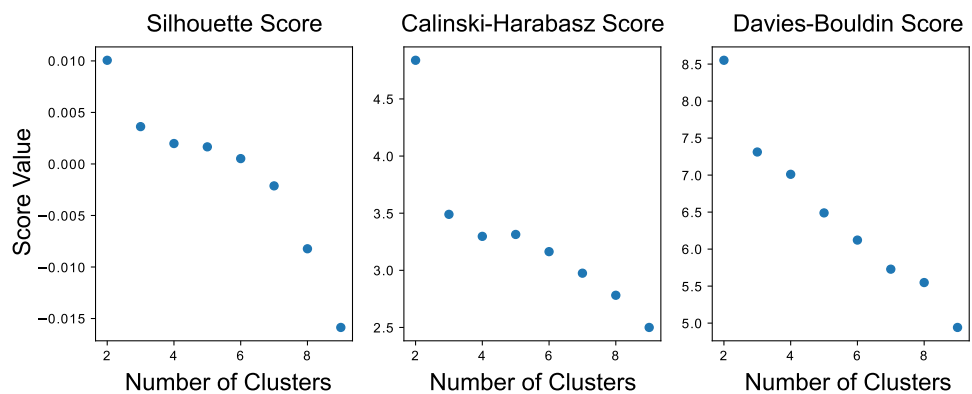

**Fig. S10**: Trp-cage cluster analysis. Cluster quality metrics for the Kmeans clustering along the separatrix.


\end{document}